# Hierarchical Stochastic Scheduling of Multi-Community Integrated Energy Systems in Uncertain Environments via Stackelberg Game

Yang Li [a,*], Bin Wang [a], Zhen Yang [b], Jiazheng Li [a], Chen Chen [c]

[a] School of Electrical Engineering, Northeast Electric Power University, Jilin 132012, China

[b] State Grid Beijing Electric Power Company, Beijing 100032, China

[c] School of Electrical Engineering, Xi'an Jiaotong University, Xi'an 710049, China

* Corresponding author. E-mail address: liyang@neepu.edu.cn (Y. Li).

**Abstract:** An operating entity utilizing community-integrated energy systems with a large number of small-scale distributed energy sources can easily trade with existing distribution markets. To solve the energy management and pricing problem of multi-community integrated energy systems (MCIESs) with multi-energy interaction, this study investigated a hierarchical stochastic optimal scheduling method for uncertain environments. To handle multiple uncertainties, a Wasserstein generative adversarial network with a gradient penalty was used to generate renewable scenarios, and the Kmeans++ clustering algorithm was employed to generate typical scenarios. A Stackelberg-based hierarchical stochastic schedule with an integrated demand response was constructed, where the MCIES operator acted as the leader pursuing the maximum net profit by setting energy prices, while the building users were followers who adjusted their energy consumption plans to minimize their total costs. Finally, a distributed iterative solution method based on a metaheuristic was designed. The effectiveness of the proposed method was verified using practical examples.

**Key Words:** Community-integrated energy systems; stochastic scheduling; Generative adversarial network; integrated demand response; Stackelberg game; renewable generation uncertainty.

## 1 Introduction

With the increasing environmental pollution and depletion of traditional fossil energy, the electric power industry is evolving toward a clean, economical, and efficient energy utilization model [1]. As potential solutions, distributed energy sources such as wind turbine (WT) and photovoltaic (PV) systems have rapidly developed. A community integrated energy system (CIES), as an operating entity that integrates a large number of distributed energy sources, can realize the complementary advantages of different energy sources [2], and easily be integrated into the existing power market. Thus, it has become a hot topic for research [3, 4].

In the context of energy marketization and power market reforms, a situation in which energy is supplied by a multi-community integrated energy system (MCIES) has gradually developed. Users and MCIES operator are different stakeholders, with prominent interactive competition characteristics [5, 6]. Reference [7] established an optimized method for operating a combined cooling, heating, and power multi-microgrid integrated energy system (IES) with electric energy interaction. Reference [8] established a multi-region IES optimization model that took into account the transmission characteristics of the district heating network (DHN). Reference [9] established several small-scale IES scheduling models to improve the overall energy utilization efficiency through energy cascade utilization technology. However, these studies mainly coordinated and optimized the energy supply subjects, while ignoring the demand response of users and the complex interest interaction between the participating subjects.

Demand Response (DR) has been considered a key measure for users to participate in the management and control process of power systems [10]. Reference [11] introduced the concept and framework of integrated demand response (IDR), which indicates the direction for the multi-energy demand side to participate in system optimization and regulation. The authors of [12] establish an IDR program considering the composition of electrical and heat loads. An IDR-based energy hub program is proposed in [13], whereby the electricity is switched to natural gas during peak hours. The above work mostly studied the optimal operation of IDR-enabled single micro energy grid or single energy hub, but did not consider the interaction of interests between different entities and the demand response potential for heat loads was not fully explored.

To characterize the interest relationship between the entities in the energy transaction process, many scholars have studied MCIESs based on game theory [14–17]. Reference [14] used cooperative games to deal with energy trading issues among multiple IESs. Reference [15] established a distributed IES hierarchical coordination scheduling method based on the Stackelberg game. Reference [16] established a bi-level collaborative optimization scheduling model for MCIESs and industrial users with different benefits. In the process of energy transaction and management, the MCIES operator first set the selling price of the energy based on the users' needs, while the users adjusted their energy demands based on the set price. There was a sequence of games between the two, and it was suitable to use the Stackelberg game model to describe the MCIES participation and interest interactions between subjects [17].

In addition, because of the increase in the penetration rate of renewable energy, how to reduce the impact of the uncertainty of wind and solar output on a system's operation has become an urgent problem to be solved [18, 19]. Up to now, some approaches such as robust optimization and stochastic programming have been utilized to handle the renewable uncertainties in IES scheduling. Robust optimization considers optimization in the worst scenarios as the core idea, which makes the generated scheduling scheme tend to be conservative [20]. As far as stochastic programming is concerned, chance-constrained programming and scenario-based method are two of the representative methods. (1) A method based on chance-constrained programming requires that the probability of establishing the constraint condition of the uncertain variable meets a certain confidence level. This method requires the use of an accurate probability density function, which is not easy to obtain [21]. (2) The scenario-based approach is another effective mathematical tool for dealing with uncertainty. This method is based on the probability distribution information of uncertain variables to sample, and uses multiple deterministic scenarios to model and solve the problem [22]. For such methods, the method used to generate a reasonable scenario is a key factor to achieve good performances.

With the development of artificial intelligence technology, the use of data-driven methods for scenario generation has attracted increasing attention. Compared with the use of traditional statistical models, deep learning methods can mine the inherent distribution of uncertain variables, solve difficult modeling problems, and realize the unsupervised generation of scenarios. Reference [23] first used generative adversarial network (GAN) to generate RG output scenarios. Reference [24] used a Wasserstein generative adversarial network with a gradient penalty (WGAN-GP) to generate a scenario with a gradient penalty.

Table 1 summarizes the main differences between the proposed model in this paper and the most relevant research studies in the field. Combining the above-mentioned existing studies, it was found that there are still some problems in MCIES energy supply and trading: (1) when an MCIES with multi-energy interaction is used to supply energy to building users, its complex interest relationship is rarely considered; (2) the demand response is considered to be a key and effective measure to stimulate the interaction between demand-side resources and renewable energy, but the demand response potential of building users has not been fully explored; and (3) there is insufficient consideration to the uncertainty of RG outputs in the process of energy optimization scheduling, which may reduce the practicability of the scheduling scheme in practice.

**Table 1 Comparison of the proposed model with the most relevant studies**

| Reference | Stakeholders | Scheduling modeling | Scenario generation | IDR | DHN transmission | Renewable Uncertainties |
|---|---|---|---|---|---|---|

|  | Upper-level | Lower-level | method | method | Electricity | Heat | characteristics | WT | PV |
|---|---|---|---|---|---|---|---|---|---|
| [5] | Energy Hubs | Resident Users | Stackelberg game | × | √ | √ | × | × | × |
| [14] | Single level, different IES as different stakeholders | | Cooperative game | × | √ | √ | √ | × | × |
| [15] | Energy dispatching center | Community operators | Stackelberg game | × | × | × | × | × | × |
| [16] | DG operators | Industrial users | Stackelberg game | × | √ | √ | √ | × | × |
| [17] | Retailer | Consumers | Stackelberg game | × | √ | × | × | × | × |
| [21] | Microgrid operator | Electric vehicle charging station | Bi-level iterative programming | × | √ | √ | × | √ | √ |
| [24] | Single level, multi-energy virtual power plant | | Two-stage robust stochastic optimal | Data-driven method: WGAN-GP | × | × | × | √ | √ |
| This paper | Multi-community integrated energy systems operator | Building users | Stackelberg game | Data-driven method: WGAN-GP | √ | √ | √ | √ | √ |

In response to these problems, this study investigated a hierarchical stochastic scheduling method based on a Stackelberg game in an uncertain environment. Compared with the existing research, the main innovations and contributions of this paper are as follows.

(1) To solve the energy management and pricing problems in the process of multiple energy transactions, a single-master multi-slave hierarchical optimization framework of a multi-energy interactive MCIES was constructed, in which, an operator of the MCIES consisting of multiple CIESs with the same interests acts as the leader and building users act as the followers.

(2) To handle the uncertainty of renewable generation outputs, a data-driven scenario analysis method was proposed by using WGAN-GP-based scenario generation, which avoided the assumption that the RG output obeyed a certain probability distribution and improved the practicability of the method.

(3) Taking into account the time delay and thermal attenuation characteristics of a district heating network, a sophisticated DHN model was established and integrated into the scheduling model. In addition, a predictive mean vote (PMV) indicator was introduced to consider the thermal comfort requirements of users in the IDR.

(4) A distributed iterative solution method based on a metaheuristic was designed. It allowed only necessary data to be transmitted between the upper and lower levels, which could effectively protect the privacy of all parties. In addition, the convergence of the algorithm was verified through an analysis of a numerical example.

## 2 Renewable scenarios generation based on WGAN-GP

GAN is an unsupervised learning model that includes two independent deep learning networks: a generator and a discriminator. The core idea of a GAN is to establish a min-max game between the generator and discriminator. In each training stage, the generator updates its weights to generate new samples, while the discriminator tries to distinguish between real historical samples and generated samples. In theory, when the game between the two reaches the Nash equilibrium, the GAN will provide a generator that can accurately reflect the distribution characteristics of the real data. Then, the discriminator cannot distinguish whether the sample comes from the generator or the historical training data. At this time, the generated sample is indistinguishable from the real historical sample. Therefore, it is as true as possible. The basic structure of a GAN is shown in Fig. 1.

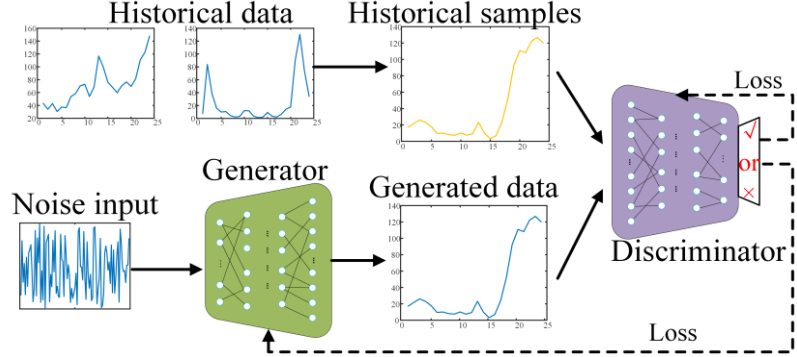

**Fig. 1 Basic structure of GAN**

Take the historical data of the RG output as the training set, and assume that the observed value of the historical sample output is *x*, and the true probability distribution is $\mathbb{P}_X$. Suppose a set of noise vectors *z* with a probability distribution $\mathbb{P}_Z$. The training goal of the generator is to make $\mathbb{P}_Z$ and $\mathbb{P}_X$ close, and the goal of the discriminator is to determine the source of the input data as closely as possible. Therefore, the loss functions of the generator and discriminator are as follows [23]:

$$L_G = \mathbb{E}_{z \sim \mathbb{P}_Z}[\log(1-D(G(z)))], \tag{1}$$

$$L_D = \mathbb{E}_{x \sim \mathbb{P}_X}[\log(D(x))] + \mathbb{E}_{z \sim \mathbb{P}_Z}[\log(1-D(G(z)))], \tag{2}$$

where $\mathbb{E}(\cdot)$ represents the expected value of the calculation; $G(z)$ denotes the sample generated by the generator; and $D(\sim)$ is the output result of the discriminator.

Then, the overall training objective function of the GAN is as follows:

$$\min_G \max_D V(D,G) = \mathbb{E}_{x \sim \mathbb{P}_X}[\log(D(x))] + \mathbb{E}_{z \sim \mathbb{P}_Z}[\log(1-D(G(z)))], \tag{3}$$

A WGAN-GP has the same structure as a GAN but uses the Wasserstein distance instead of Jensen–Shannon divergence to measure the distance between the generated data distribution and the actual data distribution. Thus, a WGAN-GP can effectively solve the problems of gradient disappearance and mode collapse found during training using the original GAN. This significantly improves the stability of the training. In addition, a WGAN-GP introduces a gradient penalty term in the discriminator, which can ensure that the discriminator approximately satisfies the 1-Lipschitz continuity. Therefore, the overall training objective function of a WGAN-GP is as follows [25]:

$$\min_G \max_D V(D,G) = \mathbb{E}_{x \sim \mathbb{P}_X}[D(x)] - \mathbb{E}_{z \sim \mathbb{P}_Z}[D(G(z))] + \lambda \mathbb{E}_{\hat{x} \sim \mathbb{P}_{\hat{x}}}[(\|\nabla_{\hat{x}} D(\hat{x})\|_2 - 1)^2], \tag{4}$$

where $\|\cdot\|_2$ denotes 2-norm; $\lambda$ is the regular term coefficient, $\hat{x} = \varepsilon x + (1-G(z))$, where here $\varepsilon \sim U[0,1]$; and *U* represents a uniform distribution.

Using the trained WGAN-GP to generate different RG output scenarios avoids pre-setting the RG output to obey a certain specific probability distribution, which has higher practicality. The trained WGAN-GP is used to generate *N* wind/photovoltaic power output scenarios, which produces a set with a total of $N^2$ scenarios. With an increase in *N*, the scenario set will grow exponentially, bringing considerable computational complexity.

Therefore, the number of scenarios needs to be reduced to reduce the amount of calculation. Clustering algorithms are usually used for scenario reduction. After using a clustering algorithm for this reduction, the probability of each scenario $\pi_k$ is calculated as follows:

$$\pi_k = \frac{N_k}{N}, \tag{5}$$

After the above steps, the numbers of WT and PV output scenarios after the reduction are $S_W$ and $S_P$, respectively. Thus, the total number of RG output scenarios is $S_{max} = S_W S_P$, and the set of RG output scenarios is $S = \{1, 2, \cdots, s, \cdots, S_{max}\}$. The corresponding probability of each scenario is $\pi_s = \pi_w \pi_p$, and the occurrence probabilities of the WT and PV output scenarios are $\pi_W$ and $\pi_P$, respectively.

## 3 Physical model of MCIES

To show the MCIES physical model, this section first introduces the overall framework of the system, and then shows how each of its equipment units were modeled.

### 3.1 Overall system framework

The structure of the MCIES studied in this work is shown in Fig. 2. The system included two CIESs, covering two forms of energy: electricity and heat. The CIESs were connected through power lines and heating pipelines, and each CIES could interact with the external power grid. The MCIES was mainly composed of WT, PV, electric boiler (EB), micro-turbine (MT), combined heat and power unit (CHP), and heat storage tank (HST) systems. It was combined with electrical energy storage (EES).

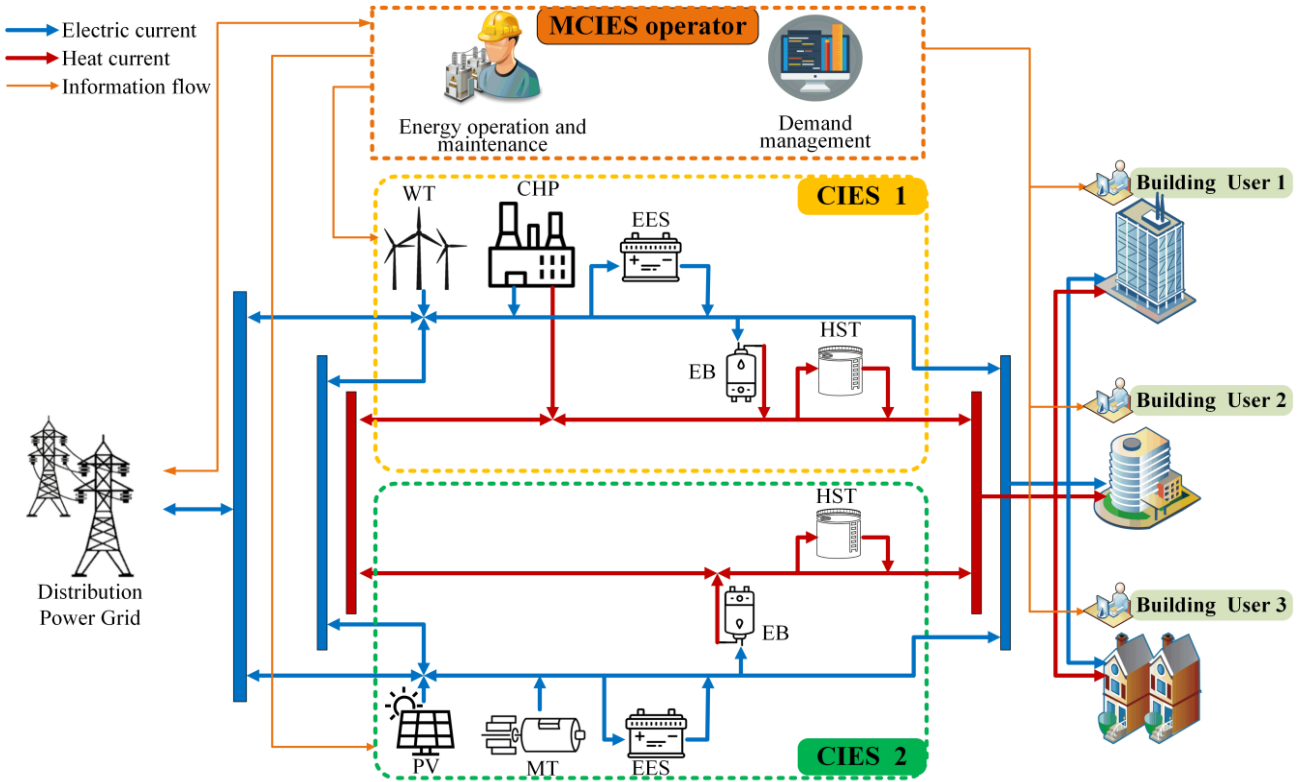

**Fig. 2 Schematic diagram of MCIES**

### 3.2 Building load model
#### 3.2.1 Electric load model

This article considers the IDR behaviors of building users. Based on the nature of the demand-side load,

the electrical load was divided into a fixed load and flexible load. The flexible load included the time-shiftable load (TSL) and interruptible load (IL) [26]. The electrical load in building $i$ during period $t$ can be described as follows:

$$P_{load,t,i} = P_{load,t,i}^0 + P_{t,i}^{TSE} - P_{t,i}^{IE}, \tag{6}$$

$$P_{t,i,\min}^{TSL} \leq P_{t,i}^{TSL} \leq P_{t,i,\max}^{TSL}, \tag{7}$$

$$\sum_{t=1}^{T} P_{t,i}^{TSL} = 0, \tag{8}$$

$$0 \leq P_{t,i}^{IL} \leq P_{t,i,\max}^{IL}, \tag{9}$$

where $P_{load,t,i}$ and $P_{load,t,i}^0$ denote the actual and initial power loads of building $i$ during period $t$, respectively; $P_{t,i}^{TSL}$ $P_{t,i,\max}^{TSL}$, and $P_{t,i,\min}^{TSL}$ are the time-shiftable load of building $i$ during period $t$, and its upper and lower limits, respectively; and $P_{t,i}^{IL}$ and $P_{t,i,\max}^{IL}$ are the interruptible load of building $i$ during period $t$ and its upper limit, respectively.

### 3.2.2 Heat load model

This study considered the thermal inertia of the building in the heat load model. Since the users' perception of a comfortable temperature has a certain flexibility, the heat load can be cuttable within an acceptable thermal comfort range for users to reduce the cost of energy. The heat load of building $i$ during period $t$ can be described as follows:

$$H_{load,t,i} = H_{load,t,i}^0 - H_{t,i}^{CH}, \tag{10}$$

where $H_{load,t,i}$ and $H_{load,t,i}^0$ denote the actual and initial thermal loads of building $i$ during period $t$, respectively; and $H_{t,i}^{CH}$ is the cuttable heat load of building $i$ during period $t$.

To quantify the acceptable thermal comfort range for users, the PMV indicator was introduced [27]:

$$PMV = 2.43 - \frac{3.76(T_s - T_{in,t})}{M(I_{cl} + 0.1)}, \tag{11}$$

where $M$ is the human energy metabolism rate; $I_{cl}$ is the thermal resistance of clothing; $T_s$ is the average temperature of human skin in a comfortable state; and $T_{in,t}$ is the indoor temperature.

According to the ISO-7730 standard, the PMV limits as follows:

$$\begin{cases} |PMV| \leq 0.9 & [1:00-7:00], [20:00-24:00] \\ |PMV| \leq 0.5 & [8:00-19:00] \end{cases}. \tag{12}$$

In addition, based on our previous work [28], the heat load in the building is calculated as follows:

$$H_{load,t,i}^0 = \frac{(T_{in,t} - T_{out,t}) + \dfrac{K_i F_i (T_{in,t} - T_{out,t})}{c_{air} \rho_{air} V_i} \Delta t}{\dfrac{1}{K_i F_i} + \dfrac{1}{c_{air} \rho_{air} V_i} \Delta t}, \tag{13}$$

where $T_{out,t}$ is the outdoor temperature during period $t$; $K_i$ is the comprehensive heat transfer coefficient of the building; $F_i$ and $V_i$ are the surface area and volume of the building, respectively; and $c_{\text{air}}$ and $\rho_{\text{air}}$ are the specific heat capacity and density of the indoor air, respectively.

To ensure user comfort, the reduction of the heat load cannot exceed its upper limit:

$$0 \leq H_{t,i}^{CH} \leq H_{load,t,i}^{0} - H_{t,i,\min}, \tag{14}$$

where $H_{t,i,\min}$ is the required minimum heat load during period $t$.

### 3.3 Electric boiler model

The output heating power of the EB should satisfy the following formulas [29]:

$$H_{t,j}^{EB} = \eta_j^{EB} P_{t,j}^{EB}, \tag{15}$$

$$0 \leq P_{t,j}^{EB} \leq P_{j,\max}^{EB}, \tag{16}$$

where $P_{t,j}^{EB}$ and $H_{t,j}^{EB}$ are the input electric power and output heat power of electric boiler $j$ during period $t$, respectively; $\eta_j^{EB}$ is the electric heating conversion coefficient of the EB; and $P_{j,\max}^{EB}$ denotes the maximum heating power of the EB.

### 3.4 CHP unit model

Based on the operating characteristics of the CHP unit, the CHP model should satisfy the following formula [30]:

$$P_{ZS,t} = P_t^{CHP} + c_v H_t^{CHP}, \tag{17}$$

$$P_{\min}^{CHP} \leq P_t^{CHP} \leq P_{\max}^{CHP}, \tag{18}$$

$$0 \leq H_t^{CHP} \leq H_{\max}^{CHP}, \tag{19}$$

$$\Delta P_{\min}^{CHP} \leq P_t^{CHP} - P_{t-1}^{CHP} \leq \Delta P_{\max}^{CHP}, \tag{20}$$

where $c_v$ is the thermoelectric ratio of a CHP; $P_t^{CHP}$ and $H_t^{CHP}$ are the output electric and heat power, respectively; $P_{ZS,t}$ is the electric energy in the condensing state; $P_{\min}^{CHP}$ and $P_{\max}^{CHP}$ are the upper and lower limits of the output electric power of a CHP, respectively; $H_{\max}^{CHP}$ is the upper limit of the output heat power of a CHP; and $\Delta P_{\max}^{CHP}$ and $\Delta P_{\min}^{CHP}$ are upper and lower limits of the climbing power, respectively.

### 3.5 MT model

The output of an MT should satisfy the following formula [31]:

$$\begin{cases} P_{\min}^{MT} \psi_t \leq P_t^{MT} \leq P_{\max}^{MT} \psi_t \\ \psi_t = 1 \ on; \ \psi_t = 0 \ off \end{cases}, \tag{21}$$

$$\Delta P_{\min}^{MT} \psi_t \leq P_t^{MT} - P_{t-1}^{MT} \leq \Delta P_{\max}^{MT} \psi_t, \tag{22}$$

where $P_t^{MT}$ is the output electric power of the MT during period $t$; $\psi_t$ is a 0/1 variable, which represents the MT state; $P_{\max}^{MT}$ and $P_{\min}^{MT}$ denote the upper and lower limits of the MT power, respectively; and $\Delta P_{\max}^{MT}$ and $\Delta P_{\min}^{MT}$ are the upper and lower limits of the MT climbing power, respectively.

### 3.6 Energy Storage Device model

The energy storage devices described in this study can be grouped into two types, namely, EES and HST. The energy storage device model is formulated as [2]:

$$C_{t+1,j} = (1-k_{loss,j})C_{t,j} + (\eta_{ch,j}P_{ch,t,j} - P_{dc,t,j}/\eta_{dc,j})\Delta t \quad \forall t, \tag{23}$$

where $C_{t+1,j}$ and $C_{t,j}$ are the capacities of energy storage device in CIES $j$ during periods $t+1$ and $t$, respectively; $P_{ch,t,j}$ and $P_{dc,t,j}$ are the charge and discharge power of energy storage device in CIES $j$, respectively; $\eta_{ch,j}$ and $\eta_{dc,j}$ are the charge and discharge efficiency coefficients, respectively; and $k_{loss,j}$ is the self-discharge rate.

In addition, the energy storage device should also meet the following constraints:

$$\begin{cases} 0 \leq P_{dc,t,j} \leq P_{dc,j,\max} \\ 0 \leq P_{ch,t,j} \leq P_{ch,j,\max} \end{cases}, \tag{24}$$

$$C_{j,\min} \leq C_{t,j} \leq C_{j,\max}, \tag{25}$$

$$C_{0,j} = C_{j,\min} = C_{T_{end},j}, \tag{26}$$

where $P_{ch,j,\max}$ and $P_{dc,j,\max}$ are the maximum charging and discharging power values of energy storage device in CIES $j$, respectively; $C_{j,\max}$ and $C_{j,\min}$ are the upper and lower limits of the capacities of energy storage device in CIES $j$, respectively; and $C_{0,j}$ and $C_{T_{end},j}$ denote the starting capacity and ending capacity of energy storage device in a scheduling period, respectively.

### 3.7 District heating network model

Generally, a DHN consists of primary and secondary heating pipe networks. The characteristics of a DHN during the heat transfer process include main two aspects: the heat loss due to heat radiation and heat delay due to the flow rate limitation [32]. Because the scale of the secondary heating pipe network is much smaller than that of the primary heating pipe network, this study ignored the secondary heating pipe network. The structure of the district heating network considered in this study is shown in Fig. 3, where H represents the heat source interaction point.

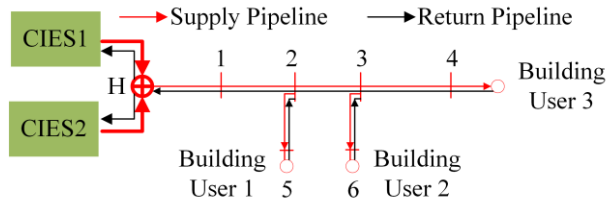

**Fig. 3 Structure diagram of district heating network**

According to the basic theorem of steady-state heat transfer, the heat loss of a heating pipe with length $L$ is calculated as follows:

$$\begin{cases} \Delta T = T_{start,t} - T_{end,t} = k_{loss}\left(T_{start,t} - T_{out,t}\right) \\ k_{loss} = 1 - e^{-\frac{\lambda L}{C_{pipe}m}} \\ \Delta H = C_{pipe}m\Delta T \end{cases}, \quad (27)$$

where $\Delta T$ is the temperature drop of the pipe; $k_{loss}$ is the temperature loss coefficient; $T_{start,t}$ and $T_{end,t}$ denote the temperatures at the head and end of the pipe during period $t$, respectively; $\lambda$ is the heat transfer efficiency per unit length of the pipe; $C_{pipe}$ is the heat capacity of the fluid; $m$ is the flow rate of the pipe; and $\Delta H$ is the heat loss.

The flow time of the medium in the tube can be approximately equal to the transmission delay time of the heating network. Thus, the heat delay time can be expressed as follows:

$$t_{delay} = \frac{\pi \rho_w L d^2}{4m}, \quad (28)$$

$$\Delta t^{sp} = round[t_{delay} / \Delta t], \quad (29)$$

where $d$ is the inner diameter of a pipe; $\rho_w$ denotes the density of water; $round[\cdot]$ is the approximation function; and $\Delta t^{sp}$ is the heat delay time of the heating pipe after approximation.

## 4 Scheduling model construction

### 4.1 MCIES scheduling model

#### 4.1.1 Objective function

The MCIES operator formulates a price strategy based on the energy side load demand, with the optimization goal of maximizing the net profit. Concerning previous research [33], this two CIESs belong to the same stakeholder and are managed by the MCIES operator. Therefore, the objective function is the sum of the net profits of two CIESs. The profit function includes three parts: the income from energy sales to users, income from interaction with the grid, and equipment operation and maintenance costs. The specific formulas for these are as follows:

$$\max\ F_M = \sum_{s=1}^{S} \pi_s \sum_{t=1}^{24} \sum_{j=1}^{J} [I_{t,s,j}^{sell} + G_{t,s,j} - C_{t,s,j}], \quad (30)$$

$$I_{t,s,j}^{sell} = \sum_{i=1}^{I}(\mu_{sell,t}P_{load,t,i} + \gamma_{sell,t}H_{load,t,i}), \quad (31)$$

$$G_{t,s,j} = (p_{sell,t}P_{t,s,j,sell}^{grid} - p_{buy,t}P_{t,s,j,buy}^{grid}), \quad (32)$$

$$\begin{aligned} C_{t,s,j} &= C^{MT} + C^{CHP} + C^{O} \\ &= [c^{MT}S_{\psi_t} + \psi_t(a^{MT} + b^{MT}P_{t,s}^{MT})] + \\ &\quad [a^{CHP}(P_{t,s}^{CHP})^2 + b^{CHP}P_{t,s}^{CHP} + c^{CHP} + \\ &\quad d^{CHP}(H_{t,s}^{CHP})^2 + e^{CHP}H_{t,s}^{CHP} + f^{CHP}P_{t,s}^{CHP}H_{t,s}^{CHP}] + \\ &\quad \sum_{e \in E_j}[P_{t,s,j,e}\beta_{s,j,e}] \end{aligned}, \quad (33)$$

where $I_{t,s,j}^{sell}$ denotes the revenue from the MCIES's energy sales to building users; $G_{t,s,j}$ is the revenue from the MCIES's interaction with the grid; $C_{t,s,j}$ is the operating cost of the MCIES; $C^{MT}$ is the MT fuel cost; $C^{CHP}$ is the CHP fuel cost; $C^{O}$ is the equipment maintenance cost; $\mu_{sell,t}$ and $\gamma_{sell,t}$ denote the MCIES's electricity and heat prices, respectively; $p_{sell,t}$ and $P_{t,s,j,sell}^{grid}$ are the electricity price and power price when the MCIES sells electricity to the grid, respectively; $p_{buy,t}$ and $P_{t,s,j,buy}^{grid}$ are the price and power when the MCIES buys electricity from the grid, respectively; $S_{\psi_t}$ and $\psi_t$ are the state variable and startup variable of an MT unit, respectively; $a^{MT}$ and $b^{MT}$ are the consumption coefficients of an MT; $c^{MT}$ is the MT start-up cost; $a^{CHP}$, $b^{CHP}$, $c^{CHP}$, $d^{CHP}$, $e^{CHP}$, and $f^{CHP}$ are the consumption coefficients of a CHP unit; $E_j$ denotes the set of all equipment in CIES $j$; $P_{t,j,e}$ denotes the output power of equipment $e$ in CIES $j$ during period $t$; and $\beta_{j,e}$ is the operation and maintenance cost factor of equipment $e$.

### 4.1.2 Electric power balance constraint

$$P_{t,s,j}^{WT} + P_{t,s,j}^{PV} + P_{t,s,j}^{CHP} + P_{t,s,j}^{MT} + P_{t,s,j}^{tl} + P_{dc,t,s,j}^{EES} + P_{t,s,j,buy}^{grid} = \sum_{i=1}^{I} P_{load,t,i} + P_{ch,t,s,j}^{EES} + P_{t,s,j}^{EB} + P_{t,s,j,sell}^{grid} \quad (34)$$

Here, $P_{t,s,j}^{WT}$ and $P_{t,s,j}^{PV}$ denote the outputs of the WT and PV systems of CIES $j$ in scenario $s$, respectively; and $P_{t,s,j}^{tl}$ is the internal transmission power of the MCIES (a positive value for $P_{t,s,j}^{tl}$ represents a power input, and a negative value represents a power output).

### 4.1.3 Thermal power balance constraint

$$H_{t,s,j}^{CHP} + H_{t,s,j}^{EB} + H_{dc,t,s,j}^{HST} + H_{t,s,j}^{tl} = \sum_{i=1}^{I}(H_{load,t+t^{sp},i} + \Delta H_{t+t^{sp},i}) + H_{ch,t,s,j}^{HST} \quad (35)$$

Here, $H_{t,s,j}^{tl}$ denotes the thermal power transmitted inside the MCIES. A positive value for $H_{t,s,j}^{tl}$ represents a power input, and a negative value represents a power output.

### 4.1.4 Internal interactive power constraints of MCIES

$$\begin{cases} \sum_{t=1}^{T} P_{t,j}^{tl} = 0 \\ P_{j,\min}^{tl} \leq P_{t,j}^{tl} \leq P_{j,\max}^{tl} \\ \sum_{t=1}^{T} H_{t,j}^{tl} = 0 \\ H_{j,\min}^{tl} \leq H_{t,j}^{tl} \leq H_{j,\max}^{tl} \end{cases} \quad (36)$$

Here, $P_{j,\max}^{tl}$ and $P_{j,\min}^{tl}$ denote the upper and lower limits of the internal interactive electric power of the MCIES, respectively; and $H_{j,\max}^{tl}$ and $H_{j,\min}^{tl}$ are the upper and lower limits of the internal interactive thermal power, respectively.

### 4.1.5 Interaction constraints between MCIES and distribution network

$$\begin{cases} P_{j,\min}^{grid} \leq P_{t,j,buy}^{grid} \leq P_{j,\max}^{grid} \\ P_{j,\min}^{grid} \leq P_{t,j,sell}^{grid} \leq P_{j,\max}^{grid} \end{cases} \quad (37)$$

Here, $P_{j,\max}^{grid}$ and $P_{j,\min}^{grid}$ are the upper and lower limits of the interactive electric power between the MCIES and the distribution network, respectively.

### 4.1.6 Real-time price constraints

According to some energy policies and to protect the interests of users, real-time electricity and heat prices should obey the following constraints:

$$\begin{cases} \mu_{\min} \leq \mu_{sell,t} \leq \mu_{\max} \\ \mu_{\min} = p_{sell,t} \\ \mu_{\max} = p_{buy,t} \\ \gamma_{\min} \leq \gamma_{sell,t} \leq \gamma_{\max} \\ \sum_{t=1}^{24} \mu_{sell,t} \leq 24\mu_{av} \\ \sum_{t=1}^{24} \gamma_{sell,t} \leq 24\gamma_{av} \end{cases}, \quad (38)$$

where $\mu_{\max}$, $\mu_{\min}$, and $\mu_{av}$ are the upper and lower limits and the average value of the electricity selling price, respectively; $\gamma_{\max}$, $\gamma_{\min}$, and $\gamma_{av}$ are the upper and lower limits and the average value of the heating price, respectively.

### 4.1.7 Equipment operation constraints

Each piece of equipment in a community needs to meet operational safety constraints. The specific constraints are shown in (15) – (29).

## 4.2 Building user model

### 4.2.1 Objective function

Building load aggregators optimize their load demand based on the energy sales price given by the MCIES, with the objective function of minimizing the total costs. Among them, the total cost of each building user mainly includes energy purchase costs and discomfort costs caused by the load response:

$$\min F_i = \sum_{t=1}^{24} \sum_{s=1}^{S} \pi_s [F_{t,i}^{\cos t} + F_{t,i}^{IDR}], \quad (39)$$

$$F_{t,i}^{\cos t} = (\mu_{sell,t} P_{load,t,i} + \gamma_{sell,t} H_{load,t,i}), \quad (40)$$

$$F_{t,i}^{IDR} = \omega_i (P_{t,i}^{TSE})^2 + \vartheta_i (P_{t,i}^{IE})^2 + \theta_i (H_{t,i}^{CH})^2, \quad (41)$$

where $F_{t,i}^{IDR}$ is the discomfort cost of building user $i$; and $\omega_i$, $\vartheta_i$, and $\theta_i$ are the discomfort cost coefficients of the time-shiftable electrical load, interruptible electrical load, and reduced heat load, respectively.

### 4.2.2 Constraints

The operating constraints on the user side of a building are detailed in (6)–(14).

## 5 Stackelberg game framework

### 5.1 Game process

The overall model framework of the single-master multiple-slave game proposed in this paper is shown in Fig. 4. As the leader, the MCIES operator maximizes their net profits by setting prices, while building users, as the followers, minimize their costs by adjusting their energy requirements. The MCIES is composed of two CIESs connected by a power line and a heat pipe. The two CIESs belong to the same stakeholder and are managed by the MCIES operator. Accordingly, the following Stackelberg game model was established:

$$G_s = \left\{ \begin{array}{c} \{MCIES \cup Users\}; \\ \{\mu_{sell,t}, \gamma_{sell,t}\}; \{P_{t,i}^{TSE}, P_{t,i}^{IE}, H_{t,i}^{CH}\}; \\ F_M; \{F_i\} \end{array} \right\}. \quad (42)$$

This game model contains three elements, namely the participants, strategy set, and benefits, which can be specifically expressed as follows.

1) Participants: $\{MCIES \cup Users\}$ represents a collection of participants, where MCIES is the leader, and Users represents a collection of building users that act as followers.

2) Strategy set: The strategy of the leader is to adjust the real-time sales prices of energy, which is expressed as $\{\mu_{sell,t}, \gamma_{sell,t}\}$; while the strategy of the followers is to continually adjust the energy consumption strategies, which are expressed as $\{P_{t,i}^{TSE}, P_{t,i}^{IE}, H_{t,i}^{CH}\}$.

3) Benefits: The benefits of each participant are their objective functions, which can be expressed as $F_M$ and $\{F_i\}$, respectively.

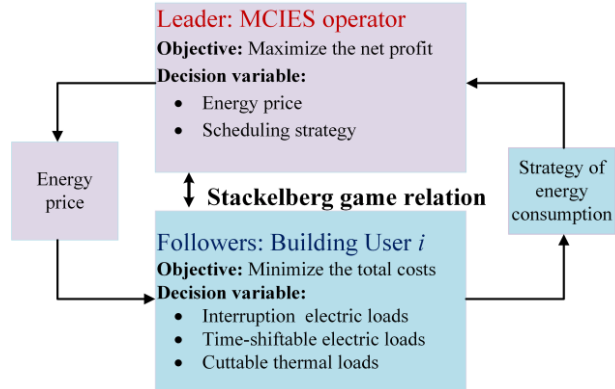

Fig. 4 Schematic diagram of Stackelberg game

### 5.2 Stackelberg equilibrium

The above-mentioned Game process will continue iteratively until the Stackelberg equilibrium (SE) is reached between different stakeholders. In this situation, the followers make an optimal response based on the leader's pricing strategy, while the leader accepts followers' optimal strategy; neither party can unilaterally change the strategy to obtain more benefits [34]. To solve the SE problem, it is necessary to prove its existence and uniqueness. The proof process is detailed in Appendix A [28, 35-40].

### 5.3 Solving algorithm

The proposed Stackelberg game model is a large-scale nonlinear problem with a bi-level structure. In general, the solutions of bi-level optimizations can be divided into two categories: 1) KKT transformation method, which uses the KKT to transform an original bi-level model into a readily solvable single-level optimization model [28]; 2) a distributed iterative solution based on an analytical method [37] or a metaheuristic algorithm [41, 42]. Although the KKT transformation method has a fast calculation speed, it has some disadvantages such as difficult conversion for large-scale nonlinear issues and poor information confidentiality. To reduce the complexity of model conversion, improve the versatility of the solution and protect the privacy of all parties, this work designs a distributed iterative solution based on a metaheuristic algorithm. The specific solution process is shown in Fig. 5. Note that here the algorithm's stopping criterion is whether the current number of iterations exceeds the pre-defined number of iterations.

Compared with the traditional centralized solution method, the designed algorithm only needs to transmit price signals and energy consumption strategies between different levels, effectively avoiding information leakage and better protecting the privacy of all the parties [5]. In this algorithm, a chaotic differential evolution in the reference [43] is used to optimize the upper MCIES operator of the proposed model. For building users in the lower level of the model, CPLEX is used to solve the problem to improve the calculation speed and accuracy. It should be highlighted that as the solution algorithm developed in this study is based on a metaheuristic, there is no guarantee that an equilibrium is always reached [16] and the algorithm's evaluation is made by simulation, which can be found in the following section.

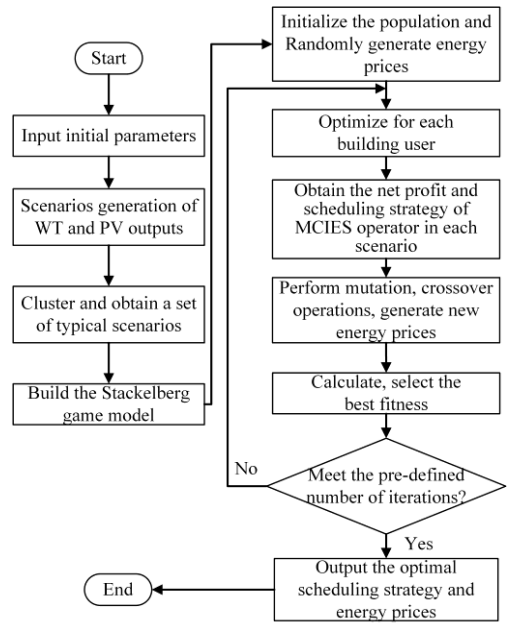

**Fig. 5 Flow chart of solution process of the Stackelberg game model**

# 6 Case study

In order to verify the feasibility of the model and method proposed in this paper, this study analyzed a multi-community-integrated energy system in North China as the research object. The TensorFlow framework was used to build the WGAN-GP. Taking the 2-year actual measured data of WT and PV in the MCIES in North China as the data set, the sampling interval of the data set is 15 minutes. In this study, 80% of the data set is selected as the training set, and 20% as the test set. The learning rate is set to 0.0002 [24, 25]. The specific parameters of the Generator and Discriminator are listed in Table 2.

The building parameters are listed in Table B1 (Appendix B). The indoor and outdoor temperature difference curve and initial heat load of each building user are shown in Fig. B1 (a) (Appendix B), and the initial electrical load is shown in Fig. B1 (b) (Appendix B). It was assumed that the time-shiftable load and interruptible load of each building user accounted for 10% of the initial electrical load at each period. The energy prices of the MCIES interacting with the distribution network and users are listed in Table 3. The parameters of the district heating network are listed in Table B2 (Appendix B), and other parameters are listed in Table B3.

**Table 2 Structure and parameters of generator and discriminator**

| Type | Layer | Title | Parameters | Numerical |
| --- | --- | --- | --- | --- |
| Generator | 1 | Fully connected layer | Number of neurons | 128 |
| | | Activation function | ReLU | — |
| | 2 | Fully connected layer | Number of neurons | 256 |
| | | Batchnorm | Dynamic mean momentum | 0.8 |
| | | Activation function | ReLU | — |
| | 3 | Fully connected layer | Number of neurons | 512 |
| | | Batchnorm | Dynamic mean momentum | 0.8 |
| | | Activation function | ReLU | — |
| | 4 | Fully connected layer | Number of neurons | 1024 |
| | | Batchnorm | Dynamic mean momentum | 0.8 |
| | | Activation function | ReLU | — |
| | 5 | Fully connected layer | Number of neurons | 24*24*1 |
| | | Activation function | tanh | — |
| Discriminator | 1 | Fully connected layer | Number of neurons | 512 |
| | | Activation function | LeakyReLU | 0.2 |
| | 2 | Fully connected layer | Number of neurons | 256 |
| | | Activation function | LeakyReLU | 0.2 |
| | 3 | Fully connected layer | Number of neurons | 1 |

**Table 3 MCIES energy price**

| Parameters | Periods/h | Value/(¥/kWh) |
|---|---|---|
| $p_{buy,t}$ (¥/kWh) | 1:00–7:00 | 0.44 |
| | 8:00–9:00, 14:00–18:00, 23:00–24:00 | 0.7 |
| | 10:00–13:00, 19:00–22:00 | 1.0 |
| $p_{sell,t}$ (¥/kWh) | 1:00–24:00 | 0.4 |
| $\gamma_{max}$ (¥/kWh) | 1:00–24:00 | 0.66 |
| $\gamma_{min}$ (¥/kWh) | 1:00–24:00 | 0.3 |
| $\mu_{av}$ (¥/kWh) | – | 0.65 |
| $\gamma_{av}$ (¥/kWh) | – | 0.5 |

**6.1 Scenario generation analysis**

**6.1.1 Model training process**

The training processes of WGAN-GP are as follows:

Step1: The random noise $z$ is inputted into the generator. In accordance with the distribution of historical samples $x$, the generator is trained to generate random samples;

Step2: The generated samples and the historical samples $x$ are fed into the discriminator simultaneously, and then the probability that the generated sample is a real sample is outputted by the discriminator;

Step3: After calculating the loss functions of the generator and the discriminator, the weights of the generator and discriminator networks are respectively updated;

Step4: The generator and discriminator are iteratively optimized until the end of training.

In Fig. 6, the change of Wasserstein distance during WGAN-GP training processes is visualized.

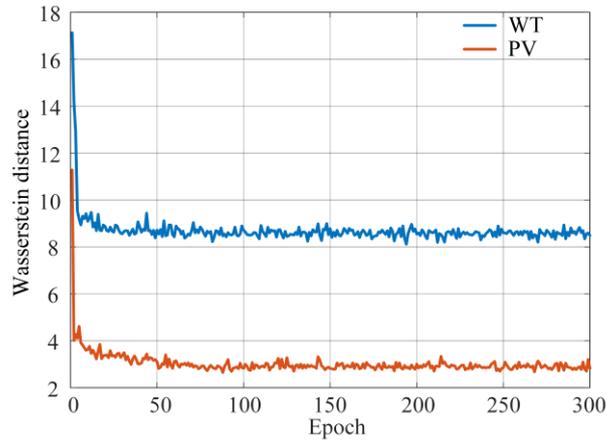

**Fig. 6 Wasserstein distance during WGAN-GP training**

From Fig.6, it can be seen that the Wasserstein distance between the real scenario and the generated scenario distribution in the WT and PV training process varied with the epoch, but in the end it gradually converged and stabilized. This shows that the WT and PV output scenarios generated by the WGAN-GP had a distribution that was very close to that of real scenarios.

**6.1.2 Cluster analysis**

To select the optimal number of clusters and clustering method, the Davies–Bouldin (DB) index was used for evaluation and analysis [44]. The DB index value was previously studied in several applications including electricity price, wind speed, and load demand [45]. Compared with other commonly-used alternative criteria such as the Silhouette coefficient, the DB index is able to yield a competitive performance with much lower complexity when verifying the clustering results [46]. Using different clustering methods, the generated WT and PV power output data were clustered into 3–10 clusters. For the hierarchical clustering method, the complete-linkage (HIA-complete) and ward-linkage (HIA-ward) methods were used for agglomerative clustering; for K-means clustering, the sample method (KM-sample), uniform scattering (KM-uniform), and Kmeans++ (KM-plus) methods were employed to initialize the cluster centroids. In addition, the Gaussian mixture model (GM) and Kmedoids method were also adopted for comparative analysis. The specific evaluation index results are shown in Fig. 7.

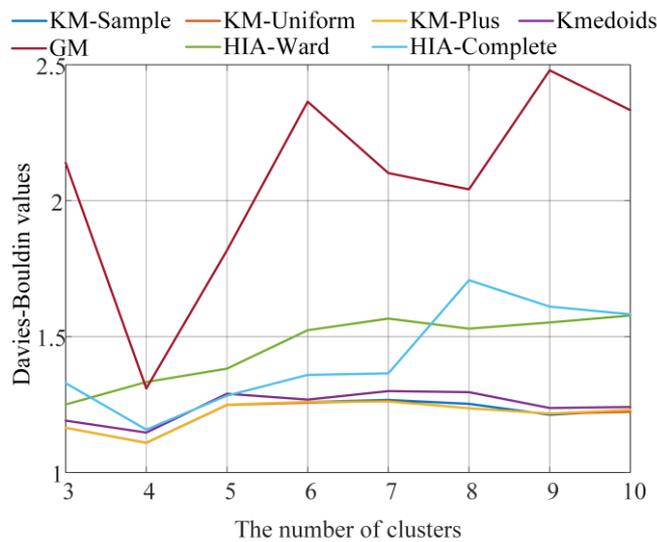

(a) Wind

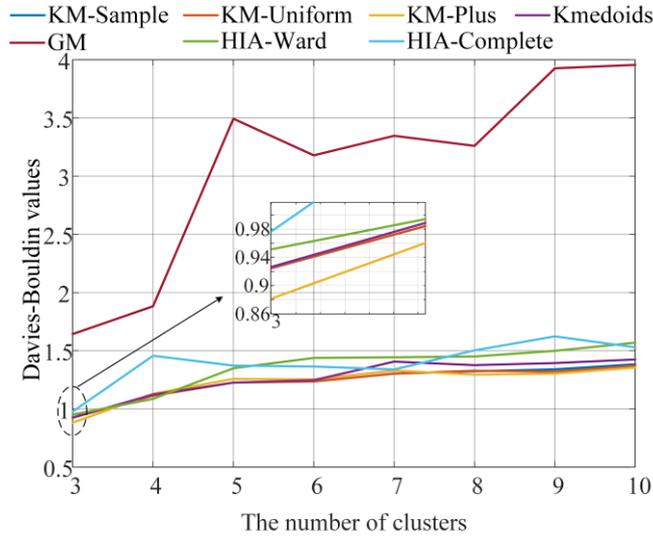

(b) Photovoltaic

**Fig. 7 Davies–Bouldin criterion values of different clustering methods and cluster numbers.**

The DB index was minimized by choosing the optimal number of clusters and clustering method. It can be seen from Fig. 7(a) that for the wind turbine output, the optimal number of clusters was R = 4, and the optimal method was Kmeans++. For the PV output, it can be observed from Fig. 7(b) that the optimal number of clusters was R = 3 and the corresponding method was Kmeans++.

**6.1.3 Temporal Correlation Analysis of Generated Scenario**

Through the above analysis, the corresponding clustering of the WT and PV output scenarios was carried out, and the results are shown in Fig. 8. The corresponding probabilities of each typical scenario after clustering are listed in Table 4. In order to examine the accuracy between the real scenario and the generated scenario, two evaluation indicators of Correlation and Normalized error are introduced for quantitative analysis. In addition, for purpose of facilitating subsequent analysis and discussion, the generated wind and solar data is extracted every 1 hour.

*1) Correlation Analysis*

The correlation of time series is capable of reflecting the real operating conditions of renewables according to temporal characteristics [47]. To properly evaluate the effectiveness of the generated scenarios, the correlation analysis has been performed in this work.

According to our previous research work [48, 49], we show the correlation coefficients between real and generated samples in the middle rows of Figs. 8(a) and 8(b). It can be known that the generated scenarios almost perfectly reproduce the characteristics of the real data while maintaining its diversity, and thereby fully representing the practical operating conditions of the WT/PV.

*2) Time Series Analysis*

The normalized errors of the generated scenarios are shown in the bottom rows of Figs. 8(a) and 8(b). Regarding WT output scenarios, the normalized error of the centroid of each generated scenario relative to that of the real data cluster is less than or equal to 0.15, and the error limits of most scenarios are not greater than 0.6; while as far as PV output scenarios are concerned, the normalized errors are much smaller than those of WT output scenarios.

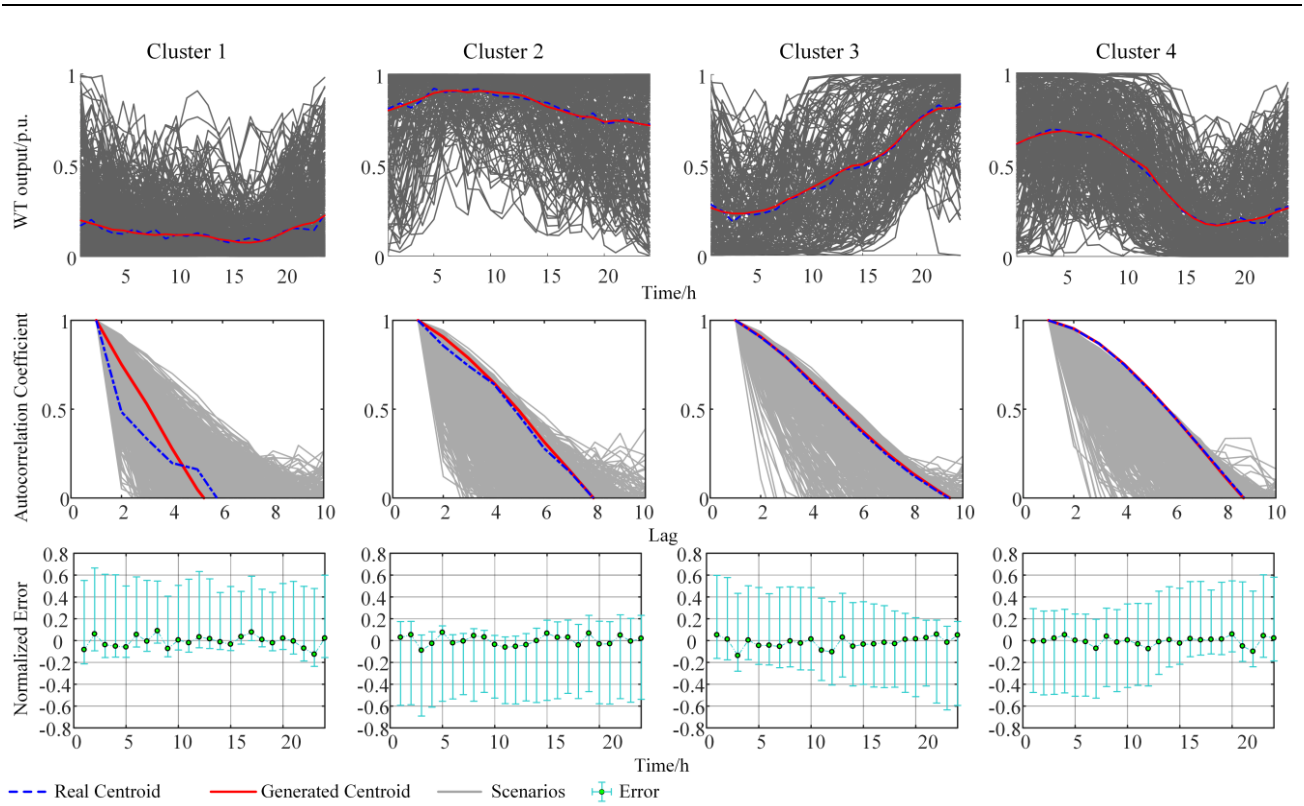

(a) Wind

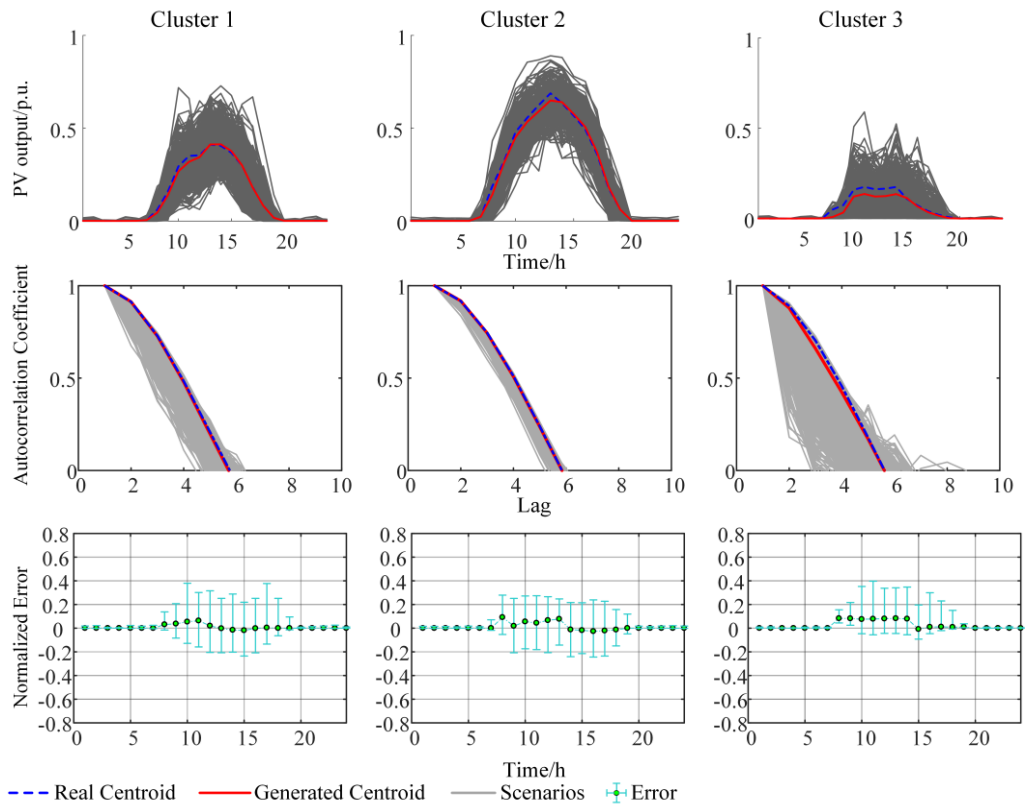

(b) Photovoltaic

**Fig. 8 Optimal clustering results under Kmeans++ method.** The top row of (a) and (b) is the comparison of the generated scenario and its centroid with the centroid of the test set. The middle row is the autocorrelation coefficient of the scenario. The bottom row is the normalized error between the generated scenario and the centroid of the test set.

Table 4 Probabilities corresponding to optimal clustering scenarios

| Type | Scenarios | Probability (%) |
|---|---|---|
| WT | Cluster 1 | 38.05 |
|    | Cluster 2 | 23.65 |
|    | Cluster 3 | 15.6 |
|    | Cluster 4 | 22.7 |
| PV | Cluster 1 | 35.05 |
|    | Cluster 2 | 26.9 |
|    | Cluster 3 | 38.05 |

## 6.2 Optimal scheduling result analysis

### 6.2.1 Simulation result

    SE could be achieved using the proposed algorithm for alternate iterations. The specific iterative optimization process for the net profit of the MCIES operator and the total cost of each building user is shown in Fig. 9.

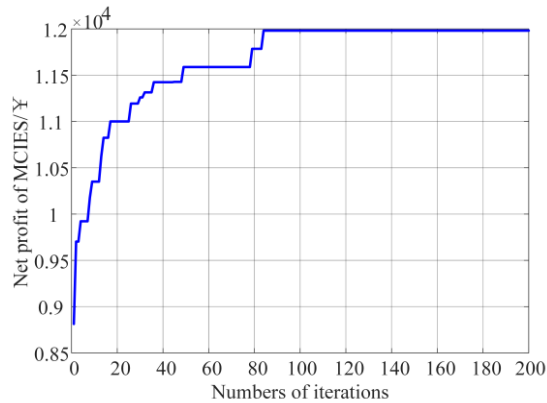

(a) Net profit of the MCIES operator

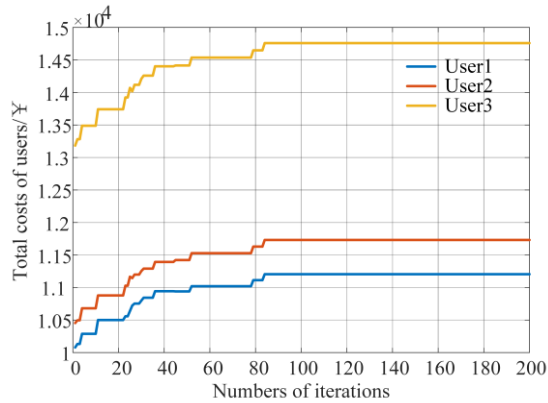

(b) Total cost of each building users

Fig. 9 Optimization iterative process of MCIES operator and each building users

    It can be seen from Fig. 9 that as the number of iterations increased, the net profit of the MCIES operator and total cost of each user of various buildings gradually increased, with convergence reached at the 83rd iteration. In other words, this was when SE was reached. In addition, it can be seen from the changing trend that the MCIES operator occupied a dominant position, and their leadership advantages would ensure that they could obtain the maximum profit, while the building users, as followers, could only make further optimal responses based on the leader's decisions.

### 6.2.2 Economic analysis

To verify the effectiveness and superiority of the method proposed in this article, the following three modes were constructed for comparative analyses.

**Mode 1: Using the proposed method, there was electricity and heating interaction in the MCIES, and considering the integrated demand response behavior of building users and the characteristics of the district heating network. Model the relationship between building users and MCIES through the Stackelberg game.**

Mode 2: There was electricity and heating interaction in the MCIES, and the characteristics of the district heating network were considered but not the user-side demand response capability. Therefore, there was no Stackelberg game relationship in this mode. The energy selling price of the MCIES remained the same as that in Mode 1.

Mode 3: There was no electricity and heating interaction in the MCIES, but the characteristics of the district heating network and the user-side demand response capability were considered. Model the relationship between building users and MCIES through the Stackelberg game.

This work compared the MCIES operator net profits and user costs of the above three Modes, and the results are listed in Table 5.

**Table 5 MCIES operator net profits and the total cost of each user**

| Mode | Net profit of MCIES operator /¥ | Total cost of each user /¥ | | |
|---|---|---|---|---|
| | | User 1 | User 2 | User 3 |
| 1 | 11981 | 11205 | 11732 | 14759 |
| 2 | 11811 | 11775 | 12291 | 15262 |
| 3 | 11756 | 11220 | 11778 | 14839 |

A comparison of Modes 1 and 2 showed that because of considering the integrated demand response, Mode 1 reduced the total cost of all users by ¥1,635 compared with Mode 2, and increased the operator's net profit by ¥170, which showed that considering the user demand response capability achieved a win-win situation on the supply and demand sides. In addition, a comparison of Modes 1 and 3 showed that considering the internal energy interaction of the MCIES further increased the operator's net profit by ¥225 while ensuring user needs. The above analysis showed that considering the integrated demand response and multi-energy interaction could further improve the overall operating economy and achieve a win-win situation for the operator and users.

#### 6.2.3 Analysis of energy interaction in the MCIES

In order to clearly demonstrate the internal interaction in the MCIES, the MCIES net profits under different interaction powers has been analyzed in this study. Here, the internal interaction energy changes in the range of [100, 700] kW with a step size of 100 kW. The change of the MCIES net profits is shown in Fig. 10.

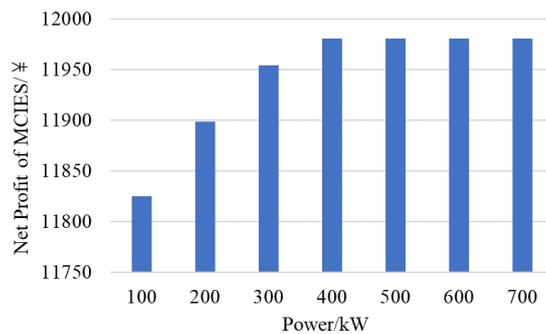

**Fig. 10 MCIES net profits under different interaction powers**

From Fig. 10, it can be observed that with the increase of the interactive power, the net profit of the MCIES gradually increases, and reaches its maximum when the interactive power is 400 kW. The reason for this is that for certain electric and heat load demands, there is an optimal solution for the intern interaction powers. Therefore, it is crucial to choose an appropriate the maximum transmission power between CIESs so as to maximize the MCIES operator's net profit.

**6.2.4 Integrated demand response analysis**

To analyze the changing trend of the load before and after the demand response on the user side, Figs.11 and 12 show the load curves for each period before and after the demand response.

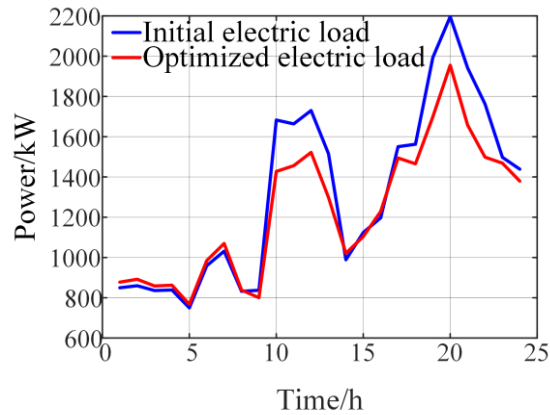

Fig. 11 Comparison of electrical load before and after demand response

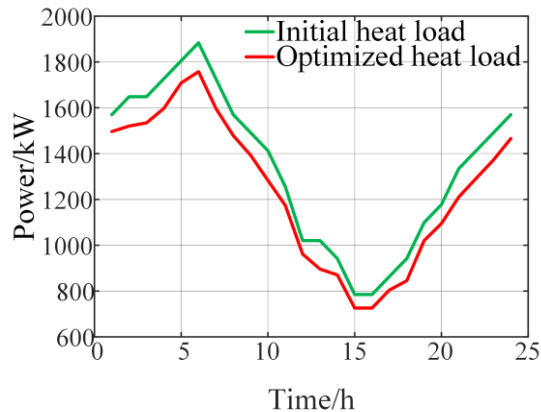

Fig. 12 Comparison of electrical load before and after demand response

From Figs.11 and 12, it can be seen that after the demand response, the electric load curve shows the characteristics of "peak shaving and valley filling," and the overall heat load is reduced. This was affected by energy prices, and building users participated in the demand response to reduce the total costs. Peaks in the initial electric load curve appeared at 10:00–13:00 and 19:00–21:00 in Fig.11. At these times, the electricity price was higher, and the user transferred and interrupted the load, which reduced the peak-to-average ratio of the electric load curve. In addition, at 10:00–15:00 in Fig.12, the users had smaller heat load reductions within the acceptable comfort range because of the low demand for heat energy by users. The results before and after each user's electric heating load demand response are recorded in Fig. B2 (Appendix B).

## 7 Conclusions

With the goal of investigating a situation involving multi-energy interaction and co-supply by a multi-community integrated energy system, this study investigated a hierarchical stochastic scheduling method based on the Stackelberg game in an uncertain environment. A WGAN-GP was used to generate renewable energy output scenarios, and clustering with Kmeans++ was used to obtain typical scenarios. With the multi-community integrated energy system operator as the leader and each building user as each follower, a single-master multi-slave hierarchical stochastic scheduling model is established, and determined a balanced interaction strategy using the proposed distributed solution algorithm. Finally, an actual calculation example verified the effectiveness of the proposed method. The main conclusions are as follows.

1) In view of the impact of renewable energy output uncertainty on optimal scheduling, this study used a data-driven scenario analysis method and a WGAN-GP to conduct an in-depth exploration of wind and solar output characteristics to generate wind and solar output scenarios, which avoided the assumption that the output obeyed a specific probability distribution, and thereby improved the practicality of the method.

2) In the established single-master multiple-slave game model, an MCIES operator guided users to adjust their energy consumption plans through energy sales prices, which achieved peak-shaving and valley-filling effects, smoothed the load curve, and reduced user costs while ensuring user comfort. In addition, through the introduction of multi-energy interaction, the economy of the MCIES was further improved.

3) A theoretical analysis proved that the proposed Stackelberg game model had a unique SE solution, which was found using the proposed distributed iterative solution method based on a metaheuristic. The results showed that the proposed algorithm had good convergence.

Note that this study did not consider the situations in which different community integrated energy systems may have competitive relationships as different stakeholders, and more than two MCIES feeding various buildings in a grid, while a more realistic scenario should take into account such situations [50]. In addition, it is a nice gap to fill to develop a novel approach for solving these situations with reducing its complexity. Another interesting topic is to investigate real-time pricing mechanisms by using machine learning techniques in smart grids [51, 52].

**Acknowledgements**

This work is supported by the Natural Science Foundation of Jilin Province, China under Grant No. YDZJ202101ZYTS149.

**Appendix A**

*(1) Proof of existence*

*Theorem:*

When the following conditions are met, there is a Stackelberg equilibrium solution [28, 35-37]:

1) The leader's objective function is the non-empty continuous function of all game strategies;

2) The objective function of each follower is a non-empty continuous function of all game strategies;

3) The objective function of each follower is the quasi-convex function of its own strategy.

*Proof:*

It is known that the net profit function of the MCIES operator ($F_M$) and the total costs of each user ($F_i$) are non-empty continuous functions of $\{\mu_{sell,t}, \gamma_{sell,t}\}$ and $\{P_{t,i}^{TSE}, P_{t,i}^{IE}, H_{t,i}^{CH}\}$. Thus, it is only necessary to prove whether condition 3) is satisfied.

According to equations (39)–(41), the second-order partial derivatives can be obtained as follows:

$$\frac{\partial^2 F_i}{\partial (P_{t,i}^{TSE})^2} = 2\omega_i > 0$$

$$\frac{\partial^2 F_i}{\partial (P_{t,i}^{IE})^2} = 2\vartheta_i > 0 \quad . \tag{1}$$

$$\frac{\partial^2 F_i}{\partial (H_{t,i}^{CH})^2} = 2\theta_i > 0$$

It can be seen from the above formula that $F_i$ is a convex function of $\{P_{t,i}^{TSE}, P_{t,i}^{IE}, H_{t,i}^{CH}\}$. In summary, the proposed game model has a Stackelberg equilibrium solution.

*(2) Proof of uniqueness*

*Theorem:*

When the game model satisfies the following conditions, there is a unique Stackelberg equilibrium solution [38-40]:

1) When the leader's strategy is given, all the followers have unique optimal solutions;

2) When the follower's strategy is given, the leader has a unique optimal solution.

*Proof:*

1) Calculate the first-order partial derivative of $F_i$ with respect to $\{P_{t,i}^{TSE}, P_{t,i}^{IE}, H_{t,i}^{CH}\}$:

$$\frac{\partial F_i}{\partial (P_{t,i}^{TSE})} = \mu_{sell,t} + 2\omega_i P_{t,i}^{TSE}$$

$$\frac{\partial F_i}{\partial (P_{t,i}^{IE})} = -\mu_{sell,t} + 2\vartheta_i P_{t,i}^{IE} \quad . \tag{2}$$

$$\frac{\partial F_i}{\partial (H_{t,i}^{CH})} = -\gamma_{sell,t} + 2\theta_i H_{t,i}^{CH}$$

Letting the above first-order partial derivatives be equal to 0 makes it possible to obtain the following:

$$P_{t,i,0}^{TSE} = -\mu_{sell,t} / (2\omega_i)$$

$$P_{t,i,0}^{IE} = \mu_{sell,t} / (2\vartheta_i) \quad . \tag{3}$$

$$H_{t,i,0}^{CH} = \gamma_{sell,t} / (2\theta_i)$$

Then, calculate the second-order partial derivative of $F_i$, with the result shown in formula (1) of the appendix. Because the discomfort cost coefficient is positive, the second-order partial derivatives are all greater than 0. Therefore, $P_{t,i,0}^{TSE}$, $P_{t,i,0}^{IE}$, and $H_{t,i,0}^{CH}$ are the minimum points of $F_i$. In addition, there are interval constraints for optimization variables. Therefore, for a given leader strategy, each follower has a unique corresponding optimal solution, and condition 1) is verified.

2) Calculate the first-order partial derivative of the MCIES operator's net profit function, $F_M$, with respect

to $\mu_{sell,t}, \gamma_{sell,t}$:

$$\frac{\partial F_M}{\partial(\mu_{sell,t})} = P_{load,t,i} > 0$$
$$\frac{\partial F_M}{\partial(\gamma_{sell,t})} = H_{load,t,i} > 0 \quad . \quad (4)$$

It can be seen from the above formula that $F_M$ is a continuously increasing function of $\mu_{sell,t}$ and $\mu_{sell,t}$. Thus, the MCIES has a unique optimal solution within the constraints of $\mu_{sell,t}$ and $\gamma_{sell,t}$.

**Appendix B**

**Table B1 Building parameters**

|  | User1 | User2 | User3 |
|---|---|---|---|
| $K$/(W·m$^{-2}$) | 0.5 | 0.5 | 0.5 |
| $F$/(m$^2$) | 4.5×10$^4$ | 5×10$^4$ | 6.2×10$^4$ |
| $V$/(m$^3$) | 4.5×10$^5$ | 5×10$^5$ | 3.72×10$^5$ |
| $C_{air}$/(kJ·kg$^{-1}$·°C$^{-1}$) | 1.007 | 1.007 | 1.007 |
| $\rho_{air}$/(kg·m$^{-3}$) | 1.2 | 1.2 | 1.2 |

**Table B2 Parameters of supply pipelines**

| Pipelines | L/(km) | d/(m) | m/(Kg/s) |
|---|---|---|---|
| H-4 | 1 | 0.6 | 200 |
| H-5 | 1.5 | 0.7 | 250 |
| H-6 | 1.8 | 0.7 | 250 |

**Table B3 MCIES operating parameters**

| Parameters | Values | Parameters | Values |
|---|---|---|---|
| $P_{j,\max}^{EB}$ (kW) | 600 | $k_{loss,j}^{EES}$ | 0.001 |
| $\eta_j^{EB}$ | 0.95 | $k_{loss,j}^{HST}$ | 0.01 |
| $P_{\max}^{CHP}$ (kW) | 1200 | $\lambda$(W/m°C) | 0.2 |
| $H_{\max}^{CHP}$ (kW) | 1200 | $C_{pipe}$ (MJ/kg°C) | 4.2×10$^{-3}$ |
| $c_v$ | 0.75 | $\rho_w$ (kg/m$^3$) | 1000 |
| $\Delta P_{\min}^{CHP}$ (kW) | -250 | $a^{MT}$ (¥/kW) | 1 |
| $\Delta P_{\max}^{CHP}$ (kW) | 250 | $b^{MT}$ (¥/kW) | 0.6 |
| $P_{\min}^{MT}$ (kW) | 50 | $c^{MT}$ (¥/kW) | 1.3 |
| $P_{\max}^{MT}$ (kW) | 500 | $a^{CHP}$ (¥/kW) | 2.415×10$^{-4}$ |
| $\Delta P_{\min}^{MT}$ (kW) | 200 | $b^{CHP}$ (¥/kW) | 0.31 |
| $\Delta P_{\max}^{MT}$ (kW) | 200 | $c^{CHP}$ (¥/kW) | 185.5 |
| $C_{1,\max}^{EES}$ (kWh) | 800 | $d^{CHP}$ (¥/kW) | 2.1×10$^{-4}$ |
| $C_{1,\min}^{EES}$ (kWh) | 100 | $e^{CHP}$ (¥/kW) | 0.0294 |
| $C_{2,\max}^{EES}$ (kWh) | 700 | $f^{CHP}$ (¥/kW) | 2.17×10$^{-7}$ |
| $C_{2,\min}^{EES}$ (kWh) | 80 | $P_{j,\min}^{tl}, H_{j,\min}^{tl}$ (kW) | -400 |
| $P_{dc,j,\max}^{EES}$ (kW) | 200 | $P_{j,\max}^{tl}, H_{j,\max}^{tl}$ (kW) | 400 |
| $P_{ch,j,\max}^{EES}$ (kW) | 200 | $P_{j,\min}^{grid}$ (kW) | -1000 |
| $\eta_{ch,j}^{EES}, \eta_{dc,j}^{EES}$ | 0.9 | $P_{j,\max}^{grid}$ (kW) | 1000 |
| $C_{j,\max}^{HST}$ (kWh) | 400 | $P_{ch,j,\max}^{HST}$ (kW) | 100 |
| $C_{j,\min}^{HST}$ (kWh) | 0 | $P_{dc,j,\max}^{HST}$ (kW) | 100 |
| $\omega_1$ (¥/kW) | 0.003 | $\theta_1$ (¥/kW) | 0.008 |
| $\omega_2$ (¥/kW) | 0.002 | $\theta_2$ (¥/kW) | 0.007 |
| $\omega_3$ (¥/kW) | 0.004 | $\theta_3$ (¥/kW) | 0.008 |
| $\vartheta_1$(¥/kW) | 0.01 | $M$(W/m$^2$) | 80 |
| $\vartheta_2$(¥/kW) | 0.013 | $I_{cl}$(m$^2$℃/W) | 0.161 |
| $\vartheta_3$(¥/kW) | 0.012 | $T_s$(℃) | 33.5 |

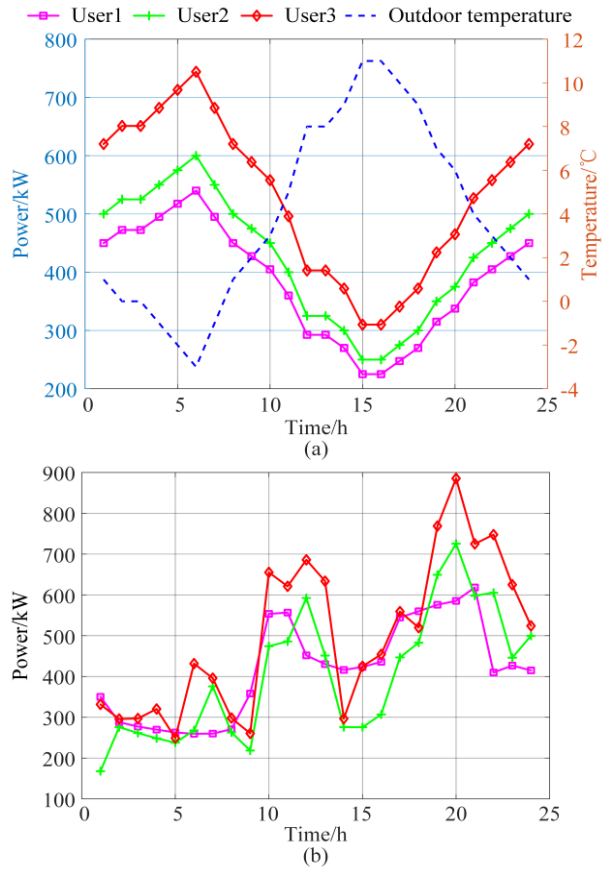

**Fig. B1 Electric load and heat load curves of building users**

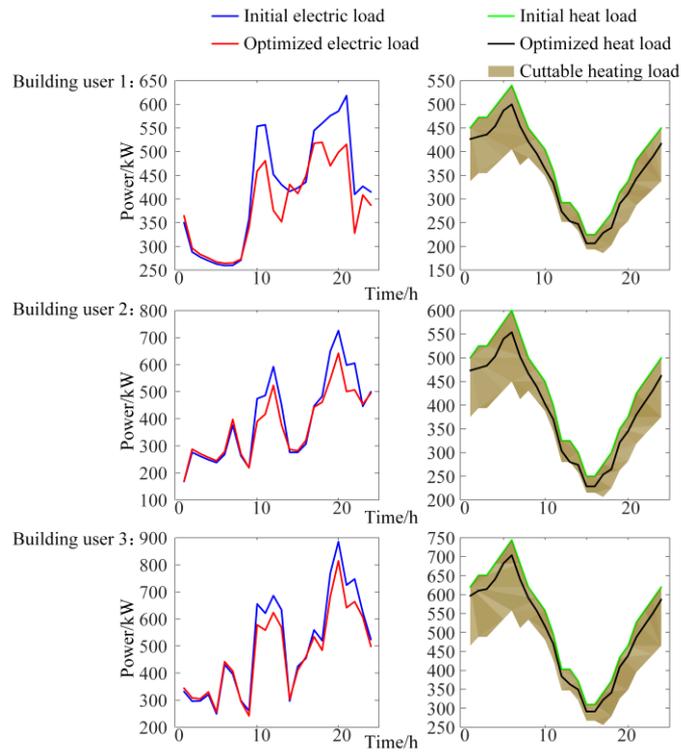

**Fig. B2 Comparison chart of results before and after optimization for each user**